# Real-time optoacoustic tracking of single moving micro-objects in deep tissue-mimicking phantoms

*Azaam Aziz[1], Mariana Medina-Sánchez[1]\*, Jing Claussen[2], and Oliver G. Schmidt[1,3]*

[1] Institute for Integrative Nanosciences, Leibniz IFW Dresden, Helmholtzstrasse 20, 01069 Dresden, Germany

[2] iThera Medical GmbH, Zielstattstrasse 13, 81379 Munich, Germany

[3] Center for Materials, Architectures and Integration of Nanomembranes (MAIN), TU Chemnitz, Reichenhainer Strasse 10, 09107 Chemnitz, Germany

\*Corresponding Author: Dr. Mariana Medina-Sánchez (m.medina.sanchez@ifw-dresden.de)

KEYWORDS: optoacoustic imaging, deep tissue imaging, micro-objects, tissue-mimicking phantom, real-time tracking, medical microbots

# ABSTRACT

Medical imaging plays an important role in diagnosis and treatment of multiple diseases. It is a field under continuous development which seeks for improved sensitivity and spatiotemporal resolution to allow the dynamic monitoring of diverse biological processes that occur at the micro- and nanoscale. Emerging technologies for targeted diagnosis and therapy such as nanotherapeutics, micro-implants, catheters and small medical tools also need to be precisely located and monitored while performing their function inside the human body. In this work, we show the real-time tracking of moving single micro-objects below centimeter thick tissue-mimicking phantoms, using multispectral optoacoustic tomography (MSOT). This technique combines the advantages of ultrasound imaging regarding depth and temporal resolution with the molecular specificity of optical methods. The resulting MSOT signal is further improved in terms of contrast and specificity by coating the micro-objects' surface with gold nanorods, exhibiting a specific absorption peak at 820 nm wavelength facilitating their discrimination from those peaks of intrinsic tissue molecules when translated to *in vivo* settings.

# INTRODUCTION

There is great interest in healthcare to perform deep tissue imaging with high spatiotemporal resolution for studying biological events like microcirculation, micro-angiogenesis, micro-lymphangiogenesis, tumor metastasis, vascularization of implanted materials, cell migration or nanotherapeutic activity in order to gain more insights into the molecular dynamics of a disease progression/regression in a non-invasive manner.[1–6] A further emerging field which is seeking for advanced imaging techniques is that of medical microrobotics. Tethered and untethered microdevices which are being developed to perform non-invasive or minimally invasive diagnosis and therapy, microsurgery,[7] and drug delivery[8,9] are usually controlled and/or guided by physical means (e.g. magnetic field, ultrasound, etc.) and need precise localization and tracking to perform the intended functional tasks.[10] So far, different imaging techniques have been explored but most of them bear insurmountable limitations. Briefly, fluorescent imaging (FI) has limited penetration depth due to significant light scattering in tissues.[11] Ultrasound (US) offers millimeter to centimeter depth penetration with micrometer resolution,[12] but it lacks providing molecular information. Computer tomography (CT) provides deep tissue penetration but utilizes ionizing radiation.[13] MRI offers submillimeter resolution and in some cases temporal resolution in the milliseconds range but demands expensive infrastructure and continuous presence of strong magnetic fields.[14] Another imaging modality, so-called PET/SPECT, which combines positron emission tomography (PET) and single photon emission computed tomography (SPECT), offers high sensitivity[15] and molecular information, but utilizes radioactive energy.

In the field of medical microrobots where real-time tracking inside deep tissue is required to bring this technology to the clinic,[10] various attempts have been carried out so far. For example, microscopic bubbles from fast-moving microjets[16] were visualized by using an ultrasound device. Servant *et al.* imaged a swimming swarm of around 10.000 artificial bacterial flagella in a living mouse by using fluorescence microscopy.[17] Likewise, Felfoul *et al.* tracked a cluster of magnetotactic bacteria ($5 \times 10^7$ bacteria per mL) by using magnetic resonance imaging.[18] Positron emission tomography (PET) combined with X-ray computed tomography (CT) was also used to image swarms of catalytic micromotors. However, this technique required radioactive isotopes and the temporal resolution (7 frames in 15 min) was low.[19] Recently, Yan *et al.* presented multifunctional bio-hybrid robots based on naturally occurring microalgae organisms, *Spirulina* *platensis* that are biodegradable, traceable by infrared and magnetic resonance and cytotoxic to cancer cells.[20] In that experiment a high concentration of material and contrast agents were needed and real-time tracking of individual or few moving micro-objects was not demonstrated.

Within this context, optoacoustic imaging appears to be a promising technique as it allows real-time imaging in deep tissue,[21] a resolution in the *µ*m range and molecular specificity which is required for distinguishing the spectral signatures of the micro-objects from surrounding tissue for future *in vivo* studies.[22–26] Multispectral optoacoustic tomography (MSOT) relies on excitation of molecules by nanosecond pulsed near-infrared laser light and detection of emitted acoustic waves by highly sensitive ultrasound detectors. When laser pulses are absorbed by tissue, the tissue expands and contracts (thermoelastic expansion), giving off pressure waves that can be detected by ultrasound detectors and accordingly mapped as 2D cross-sectional or 3D volumetric images. These images have high contrast (dependent on the molar extinction coefficient of the absorbers) and high spatial resolution (dependent on the center frequency of the transducer). This approach has been demonstrated to reach a resolution of about 150 micrometers at depths of 2–3 centimeters.[27] MSOT has been used so far to image anatomical features like human vasculature,[28,29] to assess fatty tumors,[30] monitor blood flow and oxygen state,[31] and to detect melanoma cells in sentinel lymphnodes[32] – but not yet dynamically for tracking single moving micro-objects.

Similar to other imaging modalities, contrast agents can be used to improve image contrast and targeting in MSOT. In particular nanoparticles or molecular chromophores with strong optical absorptivity and characteristic spectral profiles in the NIR spectral range are preferred. Among various contrast agents,[33,34] gold nanorods (AuNRs) are excellent candidates due to their absorption spectrum which can be tuned across the NIR region by varying their size and shape, [35,36] and their molar extinction coefficients are much higher than those of endogenous tissue chromophores like water, melanin, and hemoglobin.[37] AuNRs have been previously used for cell tracking.[38] First, AuNRs were uptaken by cells via endocytosis for further optoacoustic tracking, however, after cell uptake, the confinement and the presence of endosomes lead to the AuNRs plasmon band broaden and reduced absorbance. Therefore, the authors of this work showed that by coating the AuNRs with silica, they can provide them with steric hindrance which improve the resulting photoacoustic signal and preserved their spectral properties. This approach allowed the tracking of $2 \times 10^4$ mesenchymal stem cells in mice over a period of 15 days with high spatial resolution and without affecting the viability or differentiation potential of the transplanted cells.[38] Optoacoustic imaging has also been used to screen ovarian cancer cells for early diagnosis. In this work, AuNRs were used as passively targeted molecular imaging agents, showing the highest photo-acoustic (PA) signal ex vivo and in vivo (in mice) for an aspect ratio of 3.5. After 3h, the maximum PA signal was observed for the 2008, HEY, and SKOV3 ovarian cancer cell lines in living mice, and a linear relationship between the PA and the concentration of injected molecular imaging agent was observed with limits of detection of 0.4 nM AuNRs, being lower than other labels in the literature.[39]

In this work we present for the first time the real-time and high-precision tracking of individual micro-objects in three dimensions in tissue-mimicking phantoms down to ca. 1 cm depth. These results provide an excellent initial opportunity to demonstrate the feasibility of using MSOT to track single micro-objects in deep tissue, complying to well-accepted ethical rules according to the "3R" principles ("Replace, Reduce, Refine").[40] This means that we employ phantom tissues to avoid the early waste of living species in a research field that is still in its very infancy when considering to translate medical microrobotics into clinical practice.

# RESULTS AND DISCUSSION

## MSOT principle and setup

A cylindrical tissue-mimicking phantom with a diameter of 2 cm was placed in a holder and submerged in a water tank in a horizontal position to ensure acoustic coupling (**Figure 1A**). The laser beam and detector transducer array were in a fixed position during the entire data acquisition. The phantom was illuminated by short pulses of laser light in the near-infrared which was absorbed by the molecules inside the phantom and the embedded micro-objects. The magnitude of absorbed light was proportional to the optical absorption properties of the imaging target and local light intensity.[26] The pressure signals, resulting from thermoelastic expansion, were simultaneously collected by 256 detector elements positioned in a tomographic way and further processed to obtain cross-sectional images of the phantom in which the micro-objects were located **(Figure 1B)**. Optoacoustic signal intensities in the images correlate directly to the absorption properties of the micro-objects at the illuminated wavelengths. Images shown in this study were reconstructed using filtered back-projection algorithms.[41]

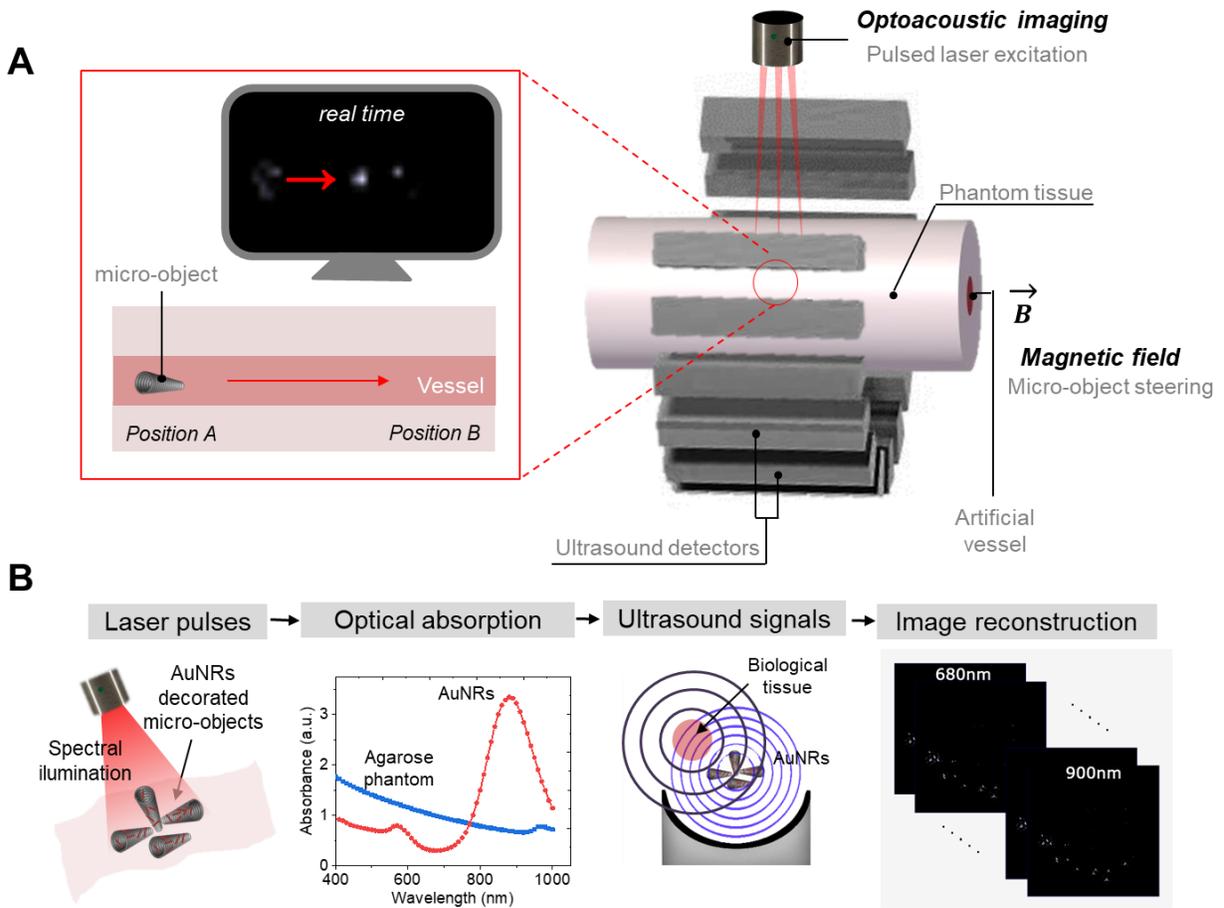

**Figure 1. Schematic of the experimental setup using multispectral optoacoustic tomography:** **(A)** MSOT imaging system, illustrating the position of the phantom, the micro-objects, the excitation source and the ultrasound detectors, **(B)** Principle of MSOT operation. The phantom is illuminated by light pulses, which then results in local heat (absorption of light) that induce pressure waves due to contraction and expansion of the molecules. These waves are captured by ultrasound detectors and sampled simultaneously using analog to digital converters. Finally, the resulting signal is converted into an optoacoustic image by using post-processing algorithms.

**Micro-objects fabrication and labeling**

3D printed conical micro-objects as model structures (diameter large opening = 12 $\mu$m, diameter small opening = 6 $\mu$m and different lengths: 25, 50 and 100 $\mu$m), were fabricated using two-photon lithography[42] and coated with Cr/Ni/Ti by electron beam evaporation to allow their magnetic propulsion and guidance (**Figure 2A**). Afterwards, the structures were covered with $Al_2O_3$ by atomic layer deposition (ALD) to improve their biocompatibility and to facilitate the surface functionalization with AuNRs. Commercially available AuNRs with an absorption peak around 820 nm and a size of 10 nm in width and 41 nm in length were covered with a dense layer of hydrophilic polymers that shield the gold surface and provide long circulation times for *in vivo* applications. The polymers also contained carboxyl groups that could be used as anchoring points for immobilization on the micro-objects. To achieve this immobilization, the $Al_2O_3$ layer of the micro-objects was first activated with oxygen plasma to increase the amount of hydroxyl groups on the surface, then immersed in (3-Aminopropyl)triethoxysilane (APTES), which binds to the hydroxyl groups through the silicon atoms of the APTES, leaving exposed amino groups. The carboxylic groups of the AuNRs were attached to these amino groups via carbodiimide chemistry. The density of the attached AuNRs was optimized by functionalizing the micro-objects with three different AuNRs concentrations (75, 150 and 300 $\mu$g/mL), which resulted in three final densities (40, 64 and 96 AuNRs/$\mu$m$^2$), after washing. SEM images in **Figure 2B** show the distribution of gold nanorods onto the micro-object surface per each case. The ratio between the MSOT signal from functionalized micro-objects (with different AuNRs concentrations) and the non-functionalized micro-objects, both previously coated with Cr/Ni/Ti/$Al_2O_3$, was obtained in the range of 680-980 nm. A well-defined absorption peak at 820 nm was observed for the different AuNRs densities. Such signal was obtained considering the average of at least three different measured objects (**Figure 2C**). The measured peak was also in good agreement with the spectrum obtained by optical spectroscopy when the initial AuNRs suspension (before functionalization) was measured (**Figure S1**).

The employed functionalization protocol is very well-stablished in the literature and has been used to attach diverse carboxylate particles or molecules such as fluorophores, enzymes, or antibodies, onto amino-modified surfaces, with good stability in physiological environments.[43] The nanostructures have been used in more complex models in vivo (mouse model) and in vitro (human cells),[39,43] preserving their stability over long periods of time and not showing cell toxicity or loses on the PA signal.

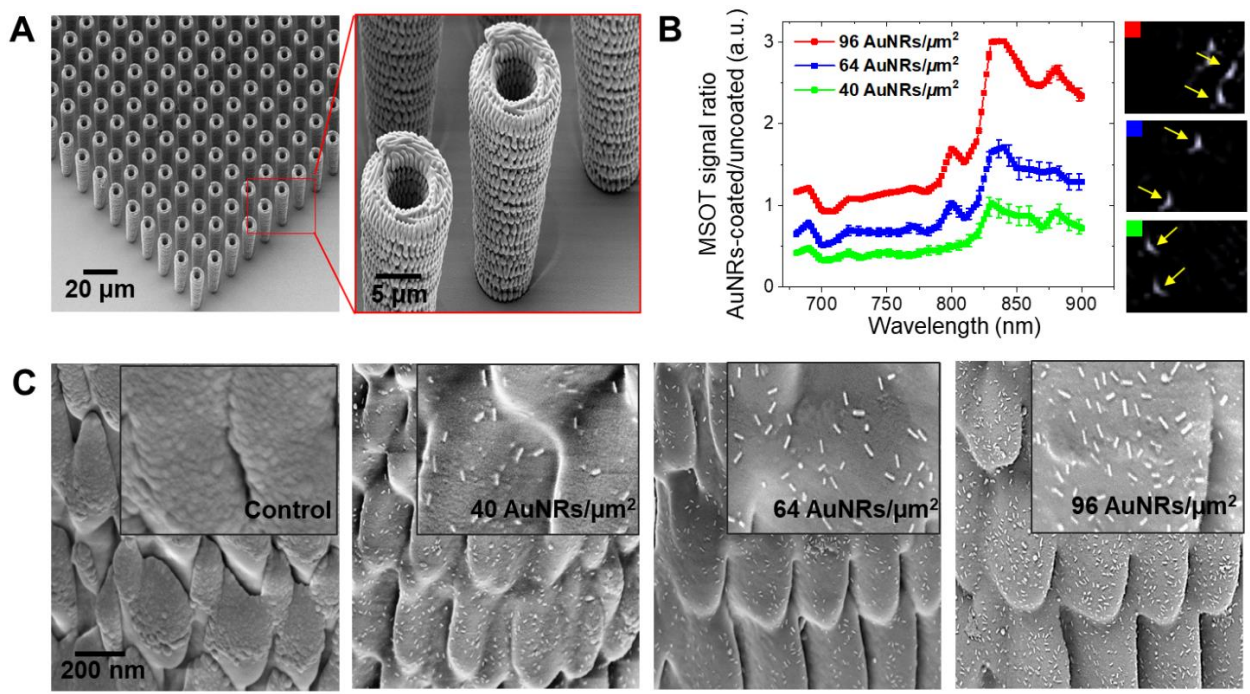

**Figure 2. Characterization of micro-objects:** (**A**) SEM images of an array of 3D printed micro-objects before AuNRs functionalization. (**B**) Spectral absorbance of the AuNRs-coated micro-objects embedded in agarose phantom for three different densities of AuNRs in relation to the non-functionalized micro-objects. The graph displays the ratio of MSOT signals obtained for AuNR-

coated micro-objects and micro-objects without AuNRs. Insets show the corresponding MSOT signals for micro-objects with three different AuNR densities. The signal corresponds to the average of three measured spots (as the ones displayed in the left side, images taken at 820 nm wavelength excitation) and the error bars correspond to the standard deviation between the different measurements. **(C)** High magnification SEM images of the micro-objects before and after functionalization with AuNRs (0, 40, 64 and 96 AuNRs/$\mu m^2$).

**Micro-objects embedded within phantoms and their corresponding MSOT signal**

Two different kinds of phantoms were fabricated, polydimethylsiloxane (PDMS)-glycerol and agarose containing soya milk, to systematically localize the micro-objects of different sizes in these two types of phantoms. The mixture of PDMS-glycerol is an emulsion formed of glycerol-filled micron-sized cavities suspended in a PDMS matrix. These phantoms are robust and easy to handle compared to agarose or gelatin-based phantoms and mimic lung and bone tissues due to their morphological similarity.[44] The absorbance of the PDMS-glycerol and agarose phantoms with a thickness of 10 mm was measured using vis-NIR spectroscopy (**Figure S2**). As the speed of sound in agarose is matched to that of the surrounding water bath, single micro-objects were visualized with high precision compared to those embedded in PDMS-glycerol phantoms which have a mismatch of 150 m/s compared to water. Agarose phantoms are also cheap, easy to produce, non-toxic, disposable and durable at high temperatures, and their properties can be easily tuned by adding different amounts of soya milk to obtain scattering coefficients similar to those of regular tissues.[45] The agarose phantom was formed in a syringe of 2 cm in diameter and 5 cm in length. A first layer of liquid agarose was poured into the syringe and solidified for 30 minutes inside a fridge. Then a drop of 10 $\mu$L of water containing the micro-objects was placed on top for drying.

Finally, a second layer of agarose was poured onto the micro-objects to complete the embedding procedure (**Figure 3A**).

In a first set of experiments, the micro-objects with a length of 100 µm were embedded within the phantoms and then visualized in single cross sections of the phantoms at a single wavelength of 820 nm (AuNRs absorption peak). All images were frequency-filtered from 1.8 MHz to 7 MHz in order to exclude the low frequency distribution from the phantom material. Each bright spot indicates the position of a single micro-object functionalized with a certain density of AuNRs (0, 40, 64 and 96 AuNRs/$\mu m^2$) (**Figure 3B, i-iv**). It was possible to locate individual micro-objects and visualize their distribution inside the phantoms. For this purpose, we took optical images of the same phantoms with micro-objects during the fabrication process before pouring the second layer of agarose. Bright-field images correlate well with the MSOT data and reveal matching locations of the micro-objects before and after imaging analysis (red rectangle, **Figure 3B, 3C and S3**). We observed that the optoacoustic signal from static micro-objects was proportional to the areal concentration of AuNRs (**Figure 3D**). Moreover, we measured the spot size, corresponding to a single micro-object, in the MSOT image finding out that it was ~5 times larger than the real length of the micro-object (100 µm) due to resolution limits imposed by the used detector array. It is also worth noting that the non-functionalized micro-objects also absorb light. This is due to the fact that the micro-objects were coated with different metal layers (Cr, Ni and Ti) before functionalization to provide them with magnetic properties and a biocompatible coating. Such materials also possesses plasmon resonance, but the intensity of the signal and its spectrum differs from the objects functionalized with gold nanorods as it can be observed in **Figure 2B and S4.**

To evaluate the accuracy of the single micro-object position, we measured the distance between micro-objects in both MSOT and bright-field images. We chose different regions of interest (ROI) where two adjacent micro-objects were located and measured the distance between them using imageJ. First, we determined the maximum pixel intensity, which corresponds to the highest part of the micro-object (large opening of the lying conical microstructure) and measured the distance between those points. The same procedure was carried out with the bright field image as shown in **Figure 3C**. The distance in MSOT (ROI1) and bright-field (ROI2) images was measured to be 770 $\mu$m and 830 $\mu$m, respectively, with a distance difference of 60 $\mu$m. The average distance difference between the micro-objects in MSOT and BF was about 108 $\mu$m, after analyzing 4 different cases (**Figure 3C and S5**). This position shift could arise due to slight displacements of micro-objects after pouring the second layer of agarose phantom.

This study was extended to smaller micro-objects (25 and 50 $\mu$m in length) and repeated under the same experimental conditions with different concentrations of AuNRs (**Figure S6**). The obtained MSOT signal depends on the concentration of AuNRs on the micro-object' surface as well as on the micro-objects size as long as it is within the spatial resolution of the used system, which is the current case (**see Figure S6**). Smaller micro-objects generate less signal as compared to the bigger micro-objects and this is due to small area covered by the same density of AuNRs. Finally, in order to confirm that MSOT is suitable to visualize other type of microstructures, we prepared agar phantoms containing a swarm of helical and cylindrical microstructures below 50 µm size (coated with Cr, Ni and Ti) and a similar MSOT signals were obtained as shown in **Figure S7**.

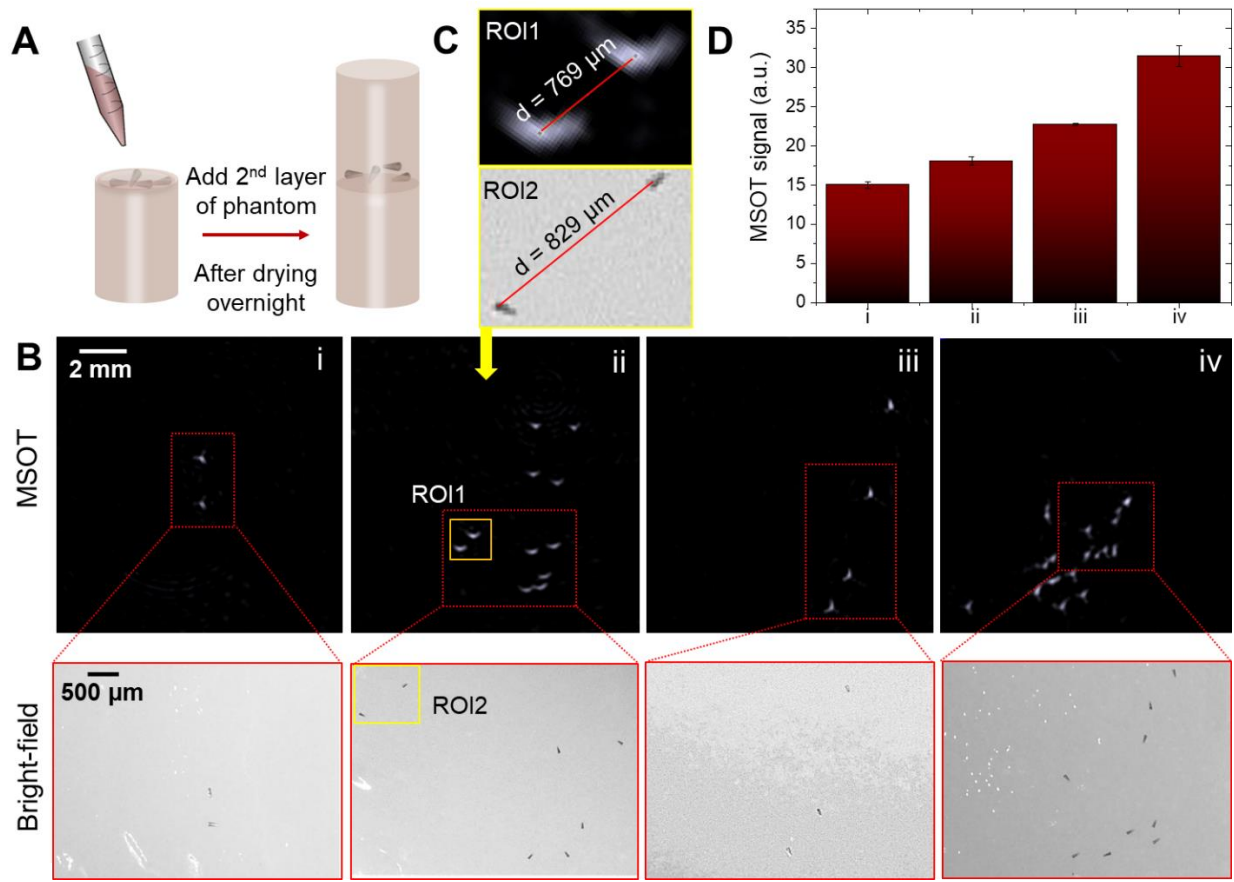

**Figure 3. MSOT signal of static micro-objects embedded within tissue mimicking phantom:** (**A**) Schematic representation of phantom preparation. Micro-objects were embedded within the phantom for static imaging studies, (**B**) MSOT signal of fully embedded micro-objects with different densities of AuNRs i) 0, ii) 40, iii) 64 and iv) 96 AuNRs/$\mu$m$^2$, and their corresponding bright-field images prior to the embedding procedure. (**C**) Comparison of the measured distance between two micro-objects derived from MSOT (770 $\mu$m) and bright field (830 $\mu$m) images, resulting in a distance difference of 60 $\mu$m, and (**D**) MSOT signals from micro-objects (100 $\mu$m in length) embedded within agarose phantoms (~ 1 cm depth) and with different density of AuNRs.

## MSOT tracking of micro-objects

In the second set of experiments, a semi-cylindrical agarose phantom with an internal channel cavity of 3 mm in diameter was prepared and placed onto a hand-held imaging detector. Acoustic gel was applied between the detector surface and the phantom for acoustic coupling (**Figure 4A**). The channel was treated with surfactant (1% w/v, SDS) to reduce the attachment of the micro-objects to the inner wall surface. Afterwards, AuNRs-coated micro-objects were injected into the channel which was later sealed to avoid undesired flows during the micro-objects' actuation. The micro-objects were moved forward by an external magnetic field (~ 60 mT). The magnetic field gradient was measured using the positioning of a magnet with respect to a Hall sensor over a distance of 0 to 3 cm (**Figure S8**). As the measurement was performed with a hand-held detector image blurring was avoided by using single pulse acquisition at a single wavelength (820 nm) with a temporal resolution of 10 frames per second (fps). Ultrasound signals were collected by 256 detector elements with an angular coverage of 270°. The data was processed by a workstation deconvoluting the signal from the electrical impulse response of the detectors. Band-pass filtering with cut-off frequencies between 1.8 and 7 MHz (maximum frequency available for the hand-held detector) was applied in order to exclude the low frequency signal distribution from the phantom. The information from MSOT images was extracted by employing image reconstruction algorithms based on the back-projection approach described elsewhere.[46] The optoacoustic imaging parameters including frequency and wavelength ranges, acquisition time, and recording speed are given in **Table 1**.

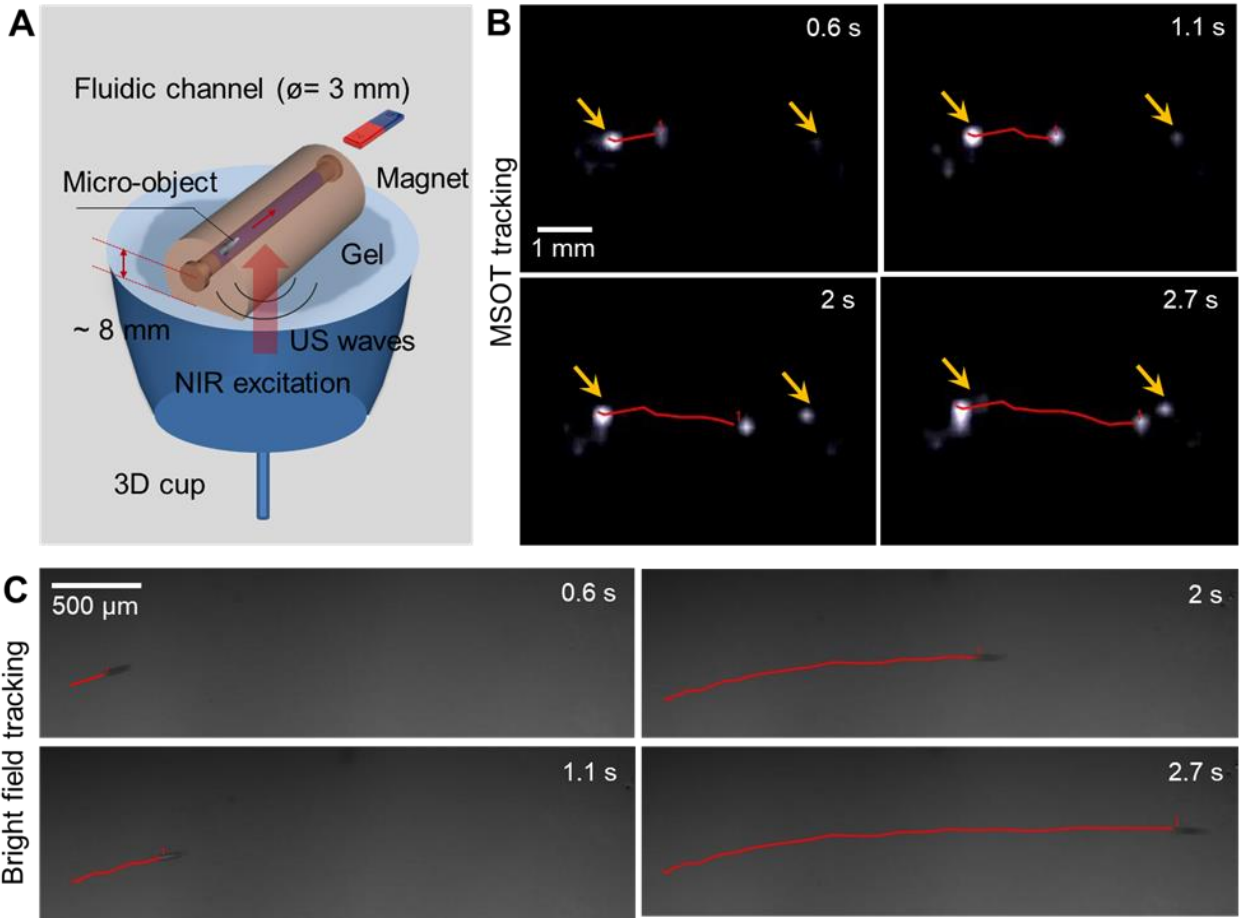

**Figure 4. Real-time tracking of single micro-object covered by 96 AuNRs/$\mu$m$^2$:** (**A**) Schematic showing the 3D hand-held detector for real time tracking of the micro-objects inside a channel cavity located approximately 8 mm deep within a tissue-mimicking phantom plus ca. 2 mm corresponding to the gel placed in between the phantom and the detector to improve the acoustic coupling, resulting in approximately 1 cm depth, (**B**) Time-lapse images of a 100 $\mu$m micro-object moving over a time period of 2.7 s (travel distance = 3.1 mm). Yellow arrows point at static micro-structures which were stuck on the phantom surface and which are used here as reference points. Red lines mark the trajectory of a single moving micro-object after different time intervals, (**C**) Optical microscopy tracking of a single micro-object over the same period of time (2.7 s) in an open agarose channel to compare the velocity and traveled distance (2.9 mm) within the same time interval.

**Table 1:** Optoacoustic imaging parameters for visualization of micro-objects in phantoms

| MSOT imaging parameters (static) | | | |
|---|---|---|---|
| **Wavelength (nm)** | 680-980 nm | **Detector (static)** | 256 elements |
| **Frequency (MHz)** | 5 MHz | **High energy** | 100 mJ |
| **Penetration depth** | 2-4 cm possible | **Spatial resolution** | 150 µm |
| **Acquisition time** | < 100 ms (single wavelength) , < 1s (multispectral) | | |
| **MSOT tracking parameters (dynamic)** | | | |
| **Wavelength (nm)** | 820 nm | **Detector (dynamic)** | 256 elements |
| **Frequency (MHz)** | 8 MHz | **Spatial resolution** | 80 µm |
| **Penetration depth** | 1.5 cm | **Frame rate (live)** | 10 fps |

**Figure 4** summarizes the results from the dynamic imaging of the micro-objects demonstrating the performance ability of MSOT. The trajectory of a single micro-object is depicted in the time-lapse images extracted from **video S1**. The single micro-object (100 µm long) took 2.7 s to travel 3.1 mm (**Figure 4B**). The speed of the AuNR-coated micro-objects based on the measured trajectories from the MSOT video was 1160 µm/s. To validate these experiments, bright-field videos were recorded in a similar fashion by preparing the same kind of phantom channel, except that the channel was left open on the top side for optical tracking (**Figure 4C).** The measured speed of the single micro-objects over 2.7 s and a travel distance of 2.9 mm was 1070 µm/s, agreeing with the results obtained from the MSOT tracking experiment. The dynamic tracking performance of a single micro-object is visualized in Video S1 comparing MSOT and bright-field tracking. The hand-held detector also allowed us to capture real-time volumetric three-dimensional images at a rate of 10 fps (**Video S2**). The average length of a moving micro-object at different locations along the trajectory was ~ 585±40 $\mu$m. To validate this, the average length of static micro-objects (imaged with different end-detector) was approximately 525±50 $\mu$m confirming similar sensitivity and spatial resolution for both detector arrays.

## CONCLUSIONS

We demonstrated that MSOT can be used to track single moving micro-objects (100 $\mu$m in length) in real-time in a tissue-mimicking phantom with a penetration depth of ~ 1 cm. Such micro-objects can be later used as drug carriers or as components of medical microrobots to perform a medical task in living organisms. We established the localization of individual micro-objects with a precision of ca. 100 $\mu$m, within phantom tissues which mimic the absorption/scattering properties of real biological tissues, as previously reported in the literature.[52]

The average length of a single micro-object under static and dynamic MSOT was determined to be 525 ± 50 $\mu$m and 585 ± 40 $\mu$m, respectively, being approximately 5 times larger than the real micro-object length (100 $\mu$m). Moreover, we could enhance the contrast and selectivity of the MSOT signal by coating the micro-objects with AuNRs as they exhibit a strong absorption peak at 820 nm wavelength, which differs from chromophores that are present in living tissues like melanin, oxyhemoglobin, water, among others, which have relatively broad absorption peaks and low absorbance.[53]

We obtained excellent agreement between the MSOT measurements and the optical analysis in both static and dynamic experiments. The ability of 3D tracking shows great potential to operate and monitor microstructures in living organisms as medical microrobots, micro-implants, microcatheters or diagnoses tools for further *in vivo* applications. In summary, MSOT emerges as a promising new tool for localization and tracking of single moving micro-objects in hard-to-reach target sites, which could significantly improve the accuracy and effectiveness of current diagnosis and therapeutics in near future.

# AUTHOR INFORMATION

**Corresponding Author**
*E-mail: m.medina.sanchez@ifw-dresden.de.
ORCID: Mariana Medina-Sanchez: 0000-0001-6149-3732

**Author Contributions**

M.M.-S., and O.G.S. conceived the project. M.M.-S. designed the experiments. A.A. performed the functionalization and characterization of the micro-objects as well as the preparation of the different phantoms. M.M.-S., A.A, and J.C. performed the experiments. All the authors analyzed and interpreted the results. A.A and M.M.-S. wrote the manuscript with contributions from J.C. M.M.-S. made the schematic figures. All authors commented and edited the manuscript and figures and approved the final version for submission.

**Notes**

The authors declare no competing financial interest.

# ACKNOLEDGEMENTS

The authors thank the priority program of the German Research Foundation SPP 1726 "Microswimmers-From Single Particle Motion to Collective Behavior" for the financial support. O.G.S. acknowledges financial support by the Leibniz Program of the German Research Foundation. We also thank L. Schwarz and F. Hebenstreit for the 3D writing of the microstructures and Dr. S. Baunack and H. Xu for the SEM images.

# REFERENCES

(1) Eriksson, S.; Nilsson, J.; Sturesson, C. *Med. Devices Evid. Res.* **2014**, *7*, 445–452.

(2) Liu, P.; Li, J.; Zhang, C.; Xu, L. X. *J. Biomed. Nanotechnol.* **2013**, *9* (6), 1041–1049.

(3) Willmann, J. K.; Paulmurugan, R.; Chen, K.; Gheysens, O.; Rodriguez-Porcel, M.; Lutz, A. M.; Chen, I. Y.; Chen, X.; Gambhir, S. S. *Radiology* **2008**, *246* (2), 508–518.

(4) Liang, C.; Diao, S.; Wang, C.; Gong, H.; Liu, T.; Hong, G.; Shi, X.; Dai, H.; Liu, Z. *Adv. Mater.* **2014**, *26* (32), 5646–5652.

(5) Liang, Y.; Li, K.; Riecken, K.; Maslyukov, A.; Gomez-Nicola, D.; Kovalchuk, Y.; Fehse, B.; Garaschuk, O. *Cell Res.* **2016**, *26* (7), 805–821.

(6) Dong-Yuan Chen, Justin Crest, and D. B. **2017**, *4* (11), 559–569.

(7) Srivastava, S. K.; Medina-Sánchez, M.; Koch, B.; Schmidt, O. G. *Adv. Mater.* **2016**, *28* (5), 832–837.

(8) Felfoul, O.; Mohammadi, M.; Taherkhani, S.; de Lanauze, D.; Xu, Y. Z.; Loghin, D.; Essa, S.; Jancik, S.; Houle, D.; Lafleur, M.; Gaboury, L.; Tabrizian, M.; Kaou, N.; Atkin,

M.; Vuong, T.; Batist, G.; Beauchemin, N.; Radzioch, D.; Martel, S. *Nat. Nanotechnol.* **2016**, *11*, 941–949.

(9) Xu, H.; Medina-Sánchez, M.; Magdanz, V.; Schwarz, L.; Hebenstreit, F.; Schmidt, O. G. *ACS Nano* **2018**, *12* (1), 327–337.

(10) Medina-Sánchez, M.; Schmidt, O. G. *Nature*. 2017, pp 406–408.

(11) Michalet X, Pinaud FF, Bentolila LA, Tsay JM, Doose S, Li JJ, Sundaresan G, Wu AM, Gambhir SS, W. S. **2008**, *538* (2005), 538–545.

(12) Bourdeau, R.; Lee-Gosselin, A.; Lakshmanan, A.; Kumar, S.; Farhadi, A.; Shapiro, M. *Nat. Publ. Gr.* **2018**, *553* (7686), 86–90.

(13) Paulus, M. J.; Gleason, S. S.; Kennel, S. J.; Hunsicker, P. R.; Johnson, D. K. *Neoplasia* **2000**, *2* (1–2), 62–70.

(14) Uecker, M.; Zhang, S.; Voit, D.; Karaus, A.; Merboldt, K. D.; Frahm, J. *NMR Biomed.* **2010**, *23* (8), 986–994.

(15) Convert, L.; Lebel, R.; Gascon, S.; Fontaine, R.; Pratte, J.-F.; Charette, P.; Aimez, V.; Lecomte, R. *J. Nucl. Med.* **2016**, *57* (9), 1460–1466.

(16) Sánchez, A.; Magdanz, V.; Schmidt, O. G.; Misra, S. In *Proceedings of the IEEE RAS and EMBS International Conference on Biomedical Robotics and Biomechatronics*; 2014; pp 169–174.

(17) Servant, A.; Qiu, F.; Mazza, M.; Kostarelos, K.; Nelson, B. J. *Adv. Mater.* **2015**, *27* (19), 2981–2988.

(18) Felfoul, O.; Mokrani, N.; Mohammadi, M.; Martel, S. In *2010 Annual International*

*Conference of the IEEE Engineering in Medicine and Biology Society, EMBC'10*; 2010.

(19)  Vilela, D.; Cossío, U.; Parmar, J.; Martínez-Villacorta, A. M.; Gómez-Vallejo, V.; Llop, J.; Sánchez, S. *ACS Nano* **2018**, *12* (2), 1220–1227.

(20)  Yan, X.; Zhou, Q.; Vincent, M.; Deng, Y.; Yu, J.; Xu, J.; Xu, T.; Tang, T.; Bian, L.; Wang, Y.-X. J.; Kostarelos, K.; Zhang, L. *Sci. Robot.* **2017**, *2* (12), eaaq1155.

(21)  Knieling, F.; Neufert, C.; Hartmann, A.; Claussen, J.; Urich, A.; Egger, C.; Vetter, M.; Fischer, S.; Pfeifer, L.; Hagel, A.; Kielisch, C.; Gortz, R. S.; Wildner, D.; Engel, M.; Rother, J.; Uter, W.; Siebler, J.; Atreya, R.; Rascher, W.; Strobel, D.; Neurath, M. F.; Waldner, M. J. *The New England journal of medicine*. United States March 2017, pp 1292–1294.

(22)  Lutzweiler, C.; Razansky, D. *Sensors (Switzerland)* **2013**, *13* (6), 7345–7384.

(23)  Mallidi, S.; Luke, G. P.; Emelianov, S. *Trends Biotecnol.* **2011**, *29* (5), 213–221.

(24)  Taruttis, A.; Ntziachristos, V. *Nat. Photonics* **2015**, *9* (4), 219–227.

(25)  Manohar, S.; Razansky, D. *Adv. Opt. Photonics* **2016**, *8* (4), 586–617.

(26)  Ntziachristos, V.; Razansky, D. **2010**, 2783–2794.

(27)  Deán-Ben, X. L.; Gottschalk, S.; Mc Larney, B.; Shoham, S.; Razansky, D. *Chem. Soc. Rev.* **2017**, *46*, 2158–2198.

(28)  Taruttis, A.; Timmermans, A. C.; Wouters, P. C.; Kacprowicz, M.; van Dam, G. M.; Ntziachristos, V. *Radiology* **2016**, *281* (1), 256–263.

(29)  Deán-Ben, X. L.; Bay, E.; Razansky, D. *Sci. Rep.* **2014**, *4*, 1–6.

(30) Buehler, A.; Diot, G.; Volz, T.; Kohlmeyer, J.; Ntziachristos, V. *J. Biophotonics* **2017**, *10* (8), 983–989.

(31) Merčep, E.; Deán-Ben, X. L.; Razansky, D. *Photoacoustics* **2018**, *10*, 48–53.

(32) Xiang, L.; Ahmad, M.; Hu, X.; Cheng, Z.; Xing, L. *X-Acoustics Imaging Sens.* **2014**, *1* (1), 18–22.

(33) Wu, D.; Huang, L.; Jiang, M. S.; Jiang, H. *Int. J. Mol. Sci.* **2014**, *15* (12), 23616–23639.

(34) Oraevsky, A. A. *Photoacoustics* **2015**, *3* (1), 1–2.

(35) Manohar, S.; Ungureanu, C.; Van Leeuwen, T. G. *Contrast Media Mol. Imaging* **2011**, *6* (5), 389–400.

(36) Hu, M.; Chen, J.; Li, Z.-Y.; Au, L.; Hartland, G. V; Li, X.; Marquez, M.; Xia, Y. *Chem. Soc. Rev.* **2006**, *35* (11), 1084–1094.

(37) Weissleder, R. *Nat. Biotechnol.* **2001**, *19* (4), 316–317.

(38) Comenge, J.; Fragueiro, O.; Sharkey, J.; Taylor, A.; Held, M.; Burton, N. C.; Park, B. K.; Wilm, B.; Murray, P.; Brust, M.; Lévy, R. *ACS Nano* **2016**, *10* (7), 7106–7116.

(39) Jokerst, J. V.; Cole, A. J.; Van De Sompel, D.; Gambhir, S. S. *ACS Nano* **2012**.

(40) No Title http://ec.europa.eu/environment/chemicals/lab_animals/3r/alternative_en.htm (accessed Jan 5, 2019).

(41) Kruger, R. A.; Reinecke, D. R.; Kruger, G. A. *Med. Phys.* **1999**, *26* (9), 1832–1837.

(42) Medina-Sánchez, M., Guix, M., Harazim, S.M., Schwarz, L., & Schmidt, O. G. *Int. Conf. Manip. Autom. Robot. Small Scales* **2016**, 1–6.

(43) Conde, J.; Ambrosone, A.; Sanz, V.; Hernandez, Y.; Marchesano, V.; Tian, F.; Child, H.; Berry, C. C.; Ibarra, M. R.; Baptista, P. V.; Tortiglione, C.; De La Fuente, J. M. *ACS Nano* **2012**, *6* (9), 8316–8324.

(44) Wróbel, M. S.; Popov, A. P.; Bykov, A. V; Tuchin, V. V. **2016**, *7* (6), 2088–2094.

(45) Zell, K.; Sperl, J. I.; Vogel, M. W.; Niessner, R.; Haisch, C. *Phys. Med. Biol.* **2007**, *52* (20), 475–484.

(46) Deán-Ben, X. L.; Razansky, D. *Opt. Express* **2013**, *21* (23), 28062–28071.

(47) Medina-Sánchez, M.; Magdanz, V.; Schwarz, L.; Xu, H.; Schmidt, O. G. In *Lecture Notes in Computer Science (including subseries Lecture Notes in Artificial Intelligence and Lecture Notes in Bioinformatics)*; 2017; Vol. 10384 LNAI, pp 579–588.

(48) Medina-Sánchez, M.; Xu, H.; Schmidt, O. G. *Ther. Deliv.* **2018**, *9* (4), 303–316.

(49) Medina-Sánchez, M.; Schwarz, L.; Meyer, A. K.; Hebenstreit, F.; Schmidt, O. G. *Nano Lett.* **2015**, *16* (1), 555–561.

(50) Magdanz, V.; Sanchez, S.; Schmidt, O. G. *Adv. Mater.* **2013**, *25* (45), 6581–6588.

(51) Xu, H.; Medina-Sánchez, M.; Magdanz, V.; Schwarz, L.; Hebenstreit, F.; Schmidt, O. G. *ACS Nano* **2017**, acsnano.7b06398.

(52) Culjat, M. O.; Goldenberg, D.; Tewari, P.; Singh, R. S. *Ultrasound Med. Biol.* **2010**, *36* (6), 861–873.

(53) Kim, S.; Chen, Y.-S.; Luke, G. P.; Emelianov, S. Y. *Biomed. Opt. Express* **2011**, *2* (9), 2540.